\documentclass{article}

\usepackage{PRIMEarxiv}

\usepackage{amsmath}
\usepackage{graphicx}
\usepackage{amsfonts}
\graphicspath{ {./figures/} }
\usepackage{hyperref}
\usepackage{float}
\usepackage{verbatim} 
\restylefloat{figure}
\usepackage{caption}
\usepackage{algorithmic}
\usepackage[]{algorithm2e}
\usepackage{relsize}
\usepackage[caption=false,font=normalsize,labelfont=sf,textfont=sf]{subfig}
\SetKwRepeat{Do}{do}{while}%

\title{Enhanced Cross-Dataset Electroencephalogram-based Emotion Recognition using Unsupervised Domain Adaptation}

\author{
  Md Niaz Imtiaz\\
  Electrical, Computer, and Biomedical Engineering \\
  Toronto Metropolitan Unversity \\
  \texttt{bguermazi@torontomu.ca} \\
  \And
  Naimul Khan\\
   Electrical, Computer, and Biomedical Engineering \\
  Toronto Metropolitan Unversity \\
  \texttt{n77khan@torontomu.ca} \\}
  
\begin{document}
  \maketitle
\begin{abstract}
Emotion recognition holds great promise in healthcare and in the development of affect-sensitive systems such as brain-computer interfaces (BCIs). However, the high cost of labeled data and significant differences in electroencephalogram (EEG) signals among individuals limit the cross-domain application of EEG-based emotion recognition models. Addressing cross-dataset scenarios poses greater challenges due to changes in subject demographics, recording devices, and stimuli presented. To tackle these challenges, we propose an improved method for classifying EEG-based emotions across domains with different distributions. We propose a Gradual Proximity-guided Target Data Selection (GPTDS) technique, which gradually selects reliable target domain samples for training based on their proximity to the source clusters and the model's confidence in predicting them. This approach avoids negative transfer caused by diverse and unreliable samples. Additionally, we introduce a cost-effective test-time augmentation (TTA) technique named Prediction Confidence-aware Test-Time Augmentation (PC-TTA). Traditional TTA methods often face substantial computational burden, limiting their practical utility. By applying TTA only when necessary, based on the model's predictive confidence, our approach improves the model's performance during inference while minimizing computational costs compared to traditional TTA approaches. Experiments on the DEAP and SEED datasets demonstrate that our method outperforms state-of-the-art approaches, achieving accuracies of 67.44\% when trained on DEAP and tested on SEED, and 59.68\% vice versa, with improvements of 7.09\% and 6.07\% over the baseline. It excels in detecting both positive and negative emotions, highlighting its effectiveness for practical emotion recognition in healthcare applications. Moreover, our proposed PC-TTA technique reduces computational time by a factor of 15 compared to traditional full TTA approaches. Code available at \url{https://github.com/RyersonMultimediaLab/EmotionRecognitionUDA}
\end{abstract}


\section{Introduction}
\label{sec:introduction}

Emotions are crucial to the human experience, affecting behavior, mental well-being, relationships, and interactions with technology \cite{kolakowska2014emotion}. Emotion recognition, a topic of growing interest, has great potential in various areas, including human-computer interaction, mood disorder management, and interactive storytelling. The implementation of accurate emotion recognition systems could lead to more natural and empathetic interactions with artificial intelligence, thereby advancing human-computer interaction.

Using physiological signals to recognize emotions is superior to relying on facial expressions, gestures, and voices, as physiological signals are less susceptible to manipulation and external influences \cite{doma2020comparative}. While multimodal approaches combining signals such as electroencephalogram (EEG), electromyography (EMG), electrocardiogram (ECG), and respiratory signals have gained interest, unimodal approaches are often preferred for their lower computational costs and simpler data collection. EEG, in particular, has proven to be a dependable and promising indicator of an individual’s mental state, as it directly captures brain activity that is challenging to manipulate \cite{li2022eeg}.

Analyzing EEG signals is a time- and labor-intensive process, making the use of existing labeled data crucial. However, substantial variations among individuals and domain shifts caused by differences in demographics, sensor technologies, and recording environments challenge traditional machine learning-based emotion recognition models, which assume identical training and test distributions. Hence, domain adaptation is a crucial task in EEG-based emotion recognition. Domain adaptation is a machine learning approach that improves a model's performance on a target domain by leveraging information from a related source domain. It aligns feature spaces or refines the model’s focus to enhance generalization to new, unlabeled data.

While considerable research has focused on domain adaptation for emotion classification using EEG, most studies have concentrated on adapting between subjects and sessions within the same dataset. Challenges persist in cross-domain scenarios, where domain adaptation across datasets is more complex due to variations in subjects, EEG collection devices, and stimuli. Limited studies on cross-dataset emotion recognition have faced performance issues largely due to these substantial differences. Our work introduces a domain-adaptive model designed to address these challenges, including domain discrepancies and the lack of labeled data in the target domain.

This method represents an advancement over our previous model \cite{imtiaz2024cross}, which was designed for predicting arrhythmia across different datasets and sessions. It consists of four stages: pre-training, cluster computation, domain adaptation, and inference. While the pre-training and cluster computation stages follow our earlier work \cite{imtiaz2024cross}, we introduce significant modifications in the domain adaptation stage with a novel technique called Gradual Proximity-guided Target Data Selection (GPTDS). Additionally, we propose a new cost-effective test-time augmentation technique, Prediction Confidence-aware Test-Time Augmentation (PC-TTA), for the inference stage.

In the pre-training stage, the model learns from labeled source samples to acquire the necessary information for emotion recognition. The cluster computation stage calculates clearly separable clusters and their centroids and other properties for the source based on true labels and for the target based on confident predictions. The adaptation stage reduces the distributional gap between the source and target domains using objective functions. In unsupervised domain adaptation, pseudo labels are used, but large distributional differences or unreliable pseudo labels can cause negative transfer \cite{liang2019exploring,zhang2022survey}. Our GPTDS approach gradually incorporates reliable target samples based on their proximity to source clusters and prediction confidence, avoiding unreliable samples that could lead to negative transfer. As training progresses and the discrepancy decreases, samples that were initially avoided due to their difficulty become eligible for selection in later stages.

Test-time augmentation (TTA) has recently gained attention for improving a model's ability to handle unseen variations and enhance classification accuracy. It works by generating multiple augmented input versions and combining their predictions. However, the computational cost of TTA is a significant concern, as it involves applying multiple transformations and performing numerous prediction operations, which can be resource-intensive. Balancing these costs with the need for high classification accuracy presents a challenge. Our PC-TTA method addresses this by quantifying predictive confidence and applying TTA only when confidence is low, reducing computational costs. While TTA has mostly been used in image classification and segmentation \cite{moshkov2020test,wang2019aleatoric,jiang2021efficientnet}, to the best of our knowledge, this is the first application of TTA in classifying EEG signals.

Our method has been evaluated on two widely used public datasets for emotion recognition: DEAP \cite{koelstra2011deap} and SEED \cite{zheng2015investigating}. This evaluation involves training on one dataset and testing on the other, in both directions. Only two previous cross-dataset studies \cite{gu2022multi,ni2021domain} have used the same training and testing settings as ours, and we compare our approach with theirs. We also contrast it with the baseline method \cite{imtiaz2024cross} and six recent high-performing domain-adaptive methods \cite{chen2021ms,li2019domain,sagawa2019distributionally,huang2020self,zhi2024confusing,jimenez2023learning}, all under identical experimental conditions. Our approach surpasses all other methods, achieving an overall accuracy of 59.68\% (SEED $\rightarrow$ DEAP) and 67.44\% (DEAP $\rightarrow$ SEED).

The key contributions of this article are as follows:

    (1) An unsupervised domain-adaptive model is proposed for emotion recognition, specifically designed to address substantial distribution differences between training and test datasets. The effectiveness of the proposed method is demonstrated through experiments, outperforming state-of-the-art approaches in cross-domain scenarios.
    
    (2) To mitigate negative transfer caused by diverse and unreliable samples, a novel technique called Gradual Proximity-guided Target Data Selection (GPTDS) is introduced. This method gradually selects reliable target domain samples for training by considering their proximity to source clusters and the model's confidence in predicting them.
    
    (3) A new, cost-effective TTA technique called Prediction Confidence-aware Test-Time Augmentation (PC-TTA) is proposed, which applies augmentation only when necessary. Experimental results show that PC-TTA significantly enhances model performance during inference and reduces the high computational costs associated with traditional TTA.

The rest of this paper is organized as follows. Section 2 discusses the existing literature relevant to our research. Section 3 provides a detailed description of the components of the proposed method, including the training and testing processes, as well as data preprocessing. Section 4 describes the datasets used and presents experiments and results for the analysis and validation of our method. In conclusion, Section 5 summarizes our work.

\section{Related work}
\label{sec:related works}

Machine learning (ML) has become essential across various domains, from human action recognition to sentiment and emotion analysis, due to its ability to handle complex, high-dimensional data \cite{kumar2023multilayer,arunnehru2022machine,revathy2022sentiment}. In EEG-based emotion recognition, support vector machines (SVMs) \cite{cai2022eeg,mazumder2019analytical,george2019recognition} are widely used. Other popular techniques include  K-Nearest Neighbor (KNN) \cite{yudhana2020human,li2018emotion}, Decision Tree (DT) \cite{jiang2019cross}, and Linear Discriminant Analysis (LDA) \cite{chen2019feature}. Shallow approaches often face challenges in effectively modeling the complex temporal relationships present in EEG signals, and they may not generalize well to new and unseen data. In the realm of deep learning, Autoencoders (AEs) \cite{zhang2020expression} and Graph Neural Networks (GNNs) \cite{lin2023eeg} have been researched. Additionally, Convolutional Neural Networks (CNNs) and Long Short-Term Memory (LSTM) networks are commonly used \cite{baradaran2023customized,jha2024emotion,ramzan2023fused,fan2024icaps}. These models have generally demonstrated strong performance, particularly in subject-dependent analyses. 

Unsupervised domain adaptation (UDA) is an effective approach for addressing distributional disparities between source and target domain data, especially when acquiring labeled data in the target domain is costly and time-consuming. It minimizes the gap by learning a mapping from source to target, as explored in research \cite{sagawa2019distributionally,huang2020self,zhi2024confusing}. Table \ref{table1} summarizes key technical insights from the literature on domain adaptation. 

\begin{table*} 

\scriptsize
\centering
\caption {Technical insights from reviewed literature.}
\begin{tabular}
{ p{0.09\linewidth} p{0.21\linewidth} p{0.15\linewidth} p{0.15\linewidth} p{0.21\linewidth}}
\hline
 Study & Approach & Advantages & Limitations/ Drawbacks & Findings and Conclusions  \\
\hline
Zhi et al. \cite{zhi2024confusing}, 2024 & Uses shared prototypes and top-2 pseudo-labels for target domain alignment, with label correction for noisy samples. & Enhances accuracy in noisy environments; clarifies classification boundaries. & Less effective with complex scenarios and many similar classes. & Outperforms with 40\% label corruption; improves accuracy by selecting 80\% prototype-nearest samples.\\
\hline
Jiménez-Guarneros and Fuentes-Pineda  \cite{jimenez2023learning}, 2023 & Multi-source Feature Alignment and Label Rectification (MFA-LR) framework for fine-grained domain alignment at subject and class levels, with pseudo-label correction. & Builds robust classifiers; maintains accuracy across data periods. & Lacks spatial relationship representation; not tested with limited target datasets. & Achieves state-of-the-art results; fine-grained alignment improves subject distribution.\\
\hline
Gu et al. \cite{gu2022multi}, 2022 & Multi-source Domain Transfer Discriminative Dictionary Learning (MDTDDL) combines transfer learning and dictionary learning with subspace projection. & Flexible learning; maintains data structure and domain correlations. & High computational complexity; limited to homogeneous domains. & Surpasses second-best methods by up to 3.30\% in accuracy; effective with small subspace dimensions and dictionary sizes.\\
\hline
He et al. \cite{he2022adversarial}, 2022 & Adversarial Discriminative Temporal Convolutional Networks (AD-TCN)  combines a temporal model with adversarial adaptive learning. & Asymmetric mapping enhances domain-specific feature extraction and representation invariance. & Faces negative transfer and distribution scatter in complex scenarios.  & Outperforms conventional methods but shows reduced effectiveness with high negative transfer.\\
\hline
Shen et al. \cite{shen2022contrastive}, 2022 & Contrastive Learning for Inter-Subject Alignment (CLISA) uses CNN-based contrastive learning to align and classify time series data from similar stimuli. & Generalizes well to new subjects; provides invariant, stimulus-generalizable representations. & Limited age group validation; may not address all shared spatiotemporal patterns. & Improves accuracy, reveals distinct neural patterns, and performs optimally with specific hyperparameters.\\
\hline
Ni et al. \cite{ni2021domain}, 2021 & Domain Adaptation Sparse Representation Classifier (DASRC) uses a domain-invariant dictionary, local information preservation, PCA, and Fisher criteria, with alternating optimization. & Enhances domain adaptation with a shared dictionary; leverages local data. & Issues with local salience integration and negative transfer. & Outperforms other methods; effective in cross-subject and cross-dataset scenarios.\\
\hline
Chen et al. \cite{chen2021ms}, 2021 & Multi-Source Marginal Distribution Adaptation (MS-MDA) uses domain-invariant and domain-specific features with separate branches per source domain. & Builds a shared dictionary and leverages both types of features. & Increased training time with more sources; needs better handling of irrelevant sources. & Outperforms other methods; effective with normalization; requires improved training efficiency.\\
\hline
Rayatdoost et al. \cite{rayatdoost2021subject}, 2021 & Combines subject-invariant learning with an adversarial network and uses a gradient reversal layer to balance recognition and subject confusion. & Reduces subject-specific biases and integrates domain adaptation. & May not fully resolve generalization issues in some cases. & Improves accuracy by reducing subject-specific biases; performs best with specific hyperparameters.\\
\hline
Huang et al. \cite{huang2020self}, 2020 & Introduces Representation Self-Challenging (RSC) to iteratively discard dominant features and use less dominant ones. & Improves cross-domain generalization; compatible with various architectures & Longer training time; requires careful hyperparameter tuning. & Effective across domains; minimal increase in model size while enhancing performance.\\
\hline
Li et al. \cite{li2019domain}, 2019 & Uses neural networks with adversarial training to adapt marginal distributions and reinforce conditional distributions. & Effective in cross-subject and cross-session scenarios; parameter-efficient. & Performance varies with feature types; requires careful tuning. & Surpasses conventional methods; better cross-session transfer.\\
\hline
Sagawa et al. \cite{sagawa2019distributionally}, 2019 & Uses Distributionally Robust Optimization (DRO) with strong regularization and a new stochastic optimizer to enhance worst-group accuracy. & Prevents spurious correlations; works with imperfect group specifications. & Requires strong regularization and a new optimizer. & Improves worst-group accuracy by 10–40 percentage points while maintaining high average accuracy.\\
\hline
\end{tabular}
\label{table1}
\end{table*}

There has been extensive research on domain adaptation for classifying EEG emotions in recent years, primarily focusing on cross-subject and cross-session adaptation \cite{chen2021ms,jimenez2023learning,guo2023multi}. Many studies emphasize feature selection to identify effective subsets from high-dimensional EEG data \cite{shen2022contrastive,she2023cross}, aiming to find common features across individuals. Adversarial learning is also popular, training models to acquire domain-invariant features using a domain discriminator to differentiate between source and target features \cite{li2019domain,huang2022generator}.

Much of the research on domain adaptation for EEG emotion recognition has focused on adapting between subjects and sessions within the same dataset. Only a few studies have explored cross-dataset scenarios \cite{gu2022multi,ni2021domain,lan2018domain,he2022adversarial,rayatdoost2021subject}. Among these, Lan et al. \cite{lan2018domain} compared existing methods on the DEAP and SEED datasets in cross-dataset settings, but they used methods originally designed for other domains and validated their model on only 3 trials from 14 subjects, rather than the full 40 trials and 32 subjects. In contrast, He et al. \cite{he2022adversarial} and Rayatdoost et al. \cite{rayatdoost2021subject} used their own self-recorded datasets. Therefore, comparisons with these methods are not feasible as they do not align with our train-test dataset settings. Instead, we compare our proposed method with the two remaining cross-dataset studies by Ni et al. \cite{ni2021domain} and Gu et al. \cite{gu2022multi}, as they conducted cross-dataset experiments using the same training and testing settings as ours.

Despite significant advancements in deep learning for EEG-based emotion recognition, challenges persist in cross-domain scenarios. While much recent research has focused on domain adaptation strategies for recognizing emotions across different subjects and sessions, limited attention has been given to the cross-dataset scenario. This scenario presents even greater challenges due to the substantial disparities between source and target domains, which arise not only from differences in subjects but also from variations in EEG recording settings and the stimuli presented. The limited studies on cross-dataset emotion recognition have struggled with performance issues, largely due to these substantial differences.

This study addresses the challenges of domain discrepancies in cross-dataset scenarios and the lack of labeled data in the target domain. We also tackle issues related to negative transfer from unreliable samples and the high computational cost of Test Time Augmentation (TTA). Although TTA has gained popularity for improving models' handling of unseen variations in tasks such as image segmentation \cite{moshkov2020test,wang2019aleatoric}, image classification \cite{kandel2021improving,jiang2021efficientnet}, and anomaly detection \cite{cohen2023boosting}, its application in the EEG domain remains unexplored. The main drawback of TTA is its high computational cost, which limits its practical use. Our PC-TTA approach mitigates this computational burden while maintaining high classification accuracy and is the first application of TTA in EEG classification.

\section{Proposed method}
\label{sec:proposed method}
\begin{figure}[!tb]
    \centering
    \includegraphics[scale=0.5]{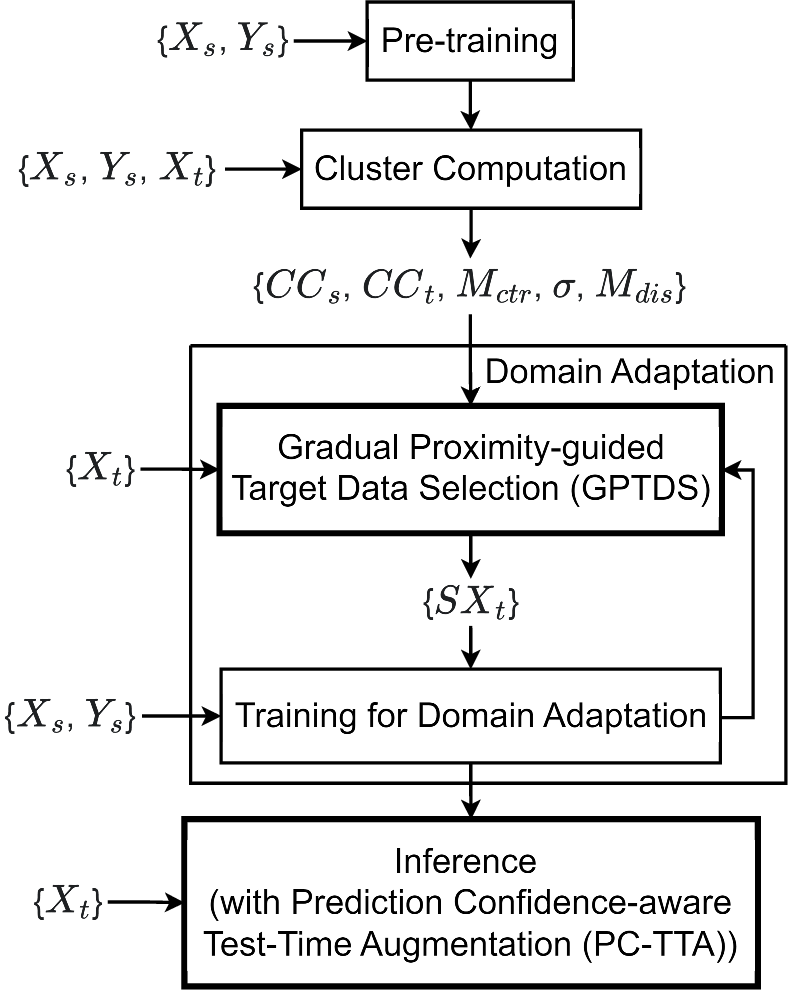}
    \caption{Block diagram illustrating the overall approach, with the two bold rectangles highlighting the new additions from this research compared to our previous model.} 
    \label{fig1}
\end{figure}

Our proposed method comprises four stages: pre-training, cluster computation, domain adaptation, and inference. While the pre-training and cluster computation stages remain the same as in our previous model \cite{imtiaz2024cross}, we introduce modifications to the domain adaptation and inference stages. Figure \ref{fig1} depicts the block diagram of the proposed approach.

\subsection{Framework}
We use \textit{source domain} to indicate the dataset utilized for training and \textit{target domain} to indicate the dataset utilized for testing. In the source domain, we have $N_s$ labeled samples,  $X_s$= $\{{x_s^i}\}_{i=1}^{N_s}$, along with their corresponding class labels, $Y_s$= $\{{y_s^i}\}_{i=1}^{N_s}$.  Conversely, in the target domain, we have $N_t$ unlabeled samples, $X_t$= $\{{x_t^i}\}_{i=1}^{N_t}$. Here, $x_s^i$ and $x_t^i$  represent the features of the $i^{th}$ EEG segment from the source and target domains, respectively (we use Power Spectral Density (PSD) as features). The marginal probability distributions of the source and target domains are $P_s(X_s)$ and $P_t(X_t)$, respectively, where $P_s(X_s)$ $\neq$ $P_t(X_t)$. Our goal is to learn a function $f$ that minimizes the gap in marginal distributions between $P_s(X_s)$ and $P_t(X_t)$, thereby enabling accurate prediction of labels for target samples.

The network architecture includes a feature extractor ($F$) followed by two parallel classifiers ($C_1$ and $C_2$) \cite{imtiaz2024cross}. Previous studies have shown that a simple feed-forward network with only fully connected layers in the feature extractor performs effectively on EEG features for emotion recognition \cite{jimenez2023learning,li2019domain}. Therefore, to keep the network architecture simple, our new feature extractor is designed with two fully connected layers, replacing the complex residual blocks used in the previous model. The batch normalization layer is incorporated to standardize features after each fully connected layer. In experiments where we used the residual block-based feature extractor from our previous model, we did not observe any performance improvements over the simpler network. Moreover, the simpler network helps mitigate the overfitting issues often encountered with more complex architectures, particularly given the relatively small datasets used in this study. We maintain two parallel classifiers following the feature extractor, as in our previous model. Having two classifiers addresses scenarios in which a single classifier may make incorrect predictions, even when the feature extractor generates distinct features. Additionally, we leverage the difference between the two classifiers to detect confident predictions in the target domain and determine the necessity of applying TTA in our PC-TTA approach. Each classifier consists of three fully connected layers. The predicted emotion category is derived by averaging the outputs from the two classifiers.

\subsubsection{Pre-training}
In the pre-training stage, the model is trained with labeled source samples ($X_s$, $Y_s$) to obtain essential information for recognizing emotions. During pre-training, we utilize a group distributionally robust optimization (DRO) technique \cite{sagawa2019distributionally}. The objective of this approach is to train models that are not reliant on misleading correlations, which can lead to poor performance on certain data groups. Instead, our goal is to train models to minimize the highest potential loss across all groups in the training data.

The loss function during the pre-training stage is the weighted sum of the \textit{classifier discrepancy loss} ($L_{dis}$), along with the \textit{classification loss} ($L_{cls}$) (\ref{eq1}). The \textit{classification loss} is computed by applying group DRO to the weighted cross-entropy loss. The \textit{classifier discrepancy loss} is calculated by evaluating the Euclidean distance between the outputs produced by the two classifiers.
\begin{align}
L= L_{cls} + \alpha  L_{dis}
\label{eq1}
\end{align}
where $\alpha$ denotes a hyperparameter.

\subsubsection{Cluster computation}
After the pre-training stage comes the cluster computation stage. Here, we start by calculating the source domain cluster centroids ($CC_{s_{pre}}$) for each emotion category in the source domain through the averaging of the feature extractor's output. Next, the model undergoes training with a pair of weighted loss functions: the \textit{cluster-separating loss} ($L_{sep}$) (\ref{eq2}) and the \textit{cluster-compacting loss} ($L_{comp}$) (\ref{eq3}), in addition to the \textit{classification loss} specified in equation (\ref{eq4}).
\begin{align}
L_{sep}= \Sigma_{k\not=l}^{K}\Sigma_{l=1}^{K} \: max(T_m-D( CC_{s_{pre}}^l, CC_{s_{pre}}^k), 0) 
\label{eq2}
\end{align}

\begin{align}
L_{comp}= \Sigma_{k=1}^{K}\Sigma_{i=1}^{n_s^k} \: D(F(x_s^i), CC_{s_{pre}}^k)
\label{eq3}
\end{align}

\begin{align}
L= L_{cls} + {\gamma}_1 L_{comp} + {\gamma}_2 L_{sep}
\label{eq4}
\end{align}

where $n_s^k$ represents the total sample count in the $k^{th}$ emotion category, $K$ represents the number of emotion categories, $D$ denotes the Euclidean distance, $T_m$ represents a large pre-defined threshold, and $\gamma_1$ and $\gamma_2$ denote hyperparameters.

Optimizing the \textit{cluster-compacting loss} reduces the intra-cluster distance, while optimizing the \textit{cluster-separating loss} increases the inter-cluster distance. After training, the clusters in the source domain become well-organized. Subsequently, we recalculate the centroids ($CC_s$) of these clusters. Since the target domain is unlabeled, we only consider confident predictions when computing the target domain clusters. We compute the \textit{mean intra-cluster distance} $M_{ctr}$ (the mean distance between the samples and their corresponding cluster centroids) (\ref{eq5}), their \textit{standard deviation} $\sigma$ (\ref{eq6}), and the \textit{mean classifier discrepancy} $M_{dis}$ (the mean disparity between the outputs of the two classifiers) (\ref{eq7}) in the source domain. We select confident predictions ($CP_t$) that meet all of the following criteria: the softmax value is greater than 0.99, the distance from the corresponding source cluster centroid is less than $M_{ctr}$, and the classifier discrepancy is less than $M_{dis}$. Afterward, we calculate the centroids ($CC_t$) of the clusters for the target domain using these selections.

\begin{align}
M_{ctr}= \frac{1}{n_s^k} \Sigma_{i=1}^{n_s^k} \: D(F(x_s^i), CC_s^k) \:for \: each \: k \in K
\label{eq5}
\end{align}
\begin{align}
\sigma= \sqrt{\frac{1}{n_s^k} \Sigma_{i=1}^{n_s^k} ((D(F(x_s^i), CC_s^k))-M_{ctr}^k)^2} \:for \: each \: k \in K
\label{eq6}
\end{align}
\begin{align}
M_{dis}= \frac{1}{N_s} \Sigma_{i=1}^{N_s} \: D(C_{1,i}, C_{2,i})
\label{eq7}
\end{align}

where $N_s$ denotes the total sample count in the source domain and $M_{ctr}^k$ represents the mean intra-cluster distance for the $k^{th}$ emotion category. 

\subsubsection{Domain adaptation}
In this proposed method for domain adaptation, we introduce improvements to our previous model. Specifically, we propose the Gradual Proximity-guided Target Data Selection (GPTDS) technique, which enhances domain adaptation compared to our previous approach. GPTDS aims to improve the selection of reliable target domain samples for training through an iterative process. In our earlier method, we selected target samples from each training batch based on their proximity to the source cluster and the model's confidence. However, samples that were distant from the source cluster and therefore excluded from training in the initial stages might still be valuable for training in later iterations as the model is trained to minimize the distributional discrepancy between the source and target domains in the domain adaptation stage. Our previous method did not account for this, leading to a low number of samples included in the training process, overlooking potentially eligible candidates for later stages. GPTDS addresses this issue. 

We first calculate the feature maps $F(X_t)$ by inputting the target domain samples $X_t$ into the pre-trained network. Next, we determine the similarity of the target domain samples to the source samples by calculating the distance between the feature map of a target sample and the centroid of the corresponding source cluster. We select candidates ($CX_t$) for training from the target domain that are more similar to the source samples, based on the following condition:

\begin{align}
CX_t= \{x_t\in X_t \mid D(F(x_t), CC_s^k) < (M_{ctr}^k + \sigma^k/2)\}
\label{eq8}
\end{align}

where $CC_s^k$ denotes the cluster centroid, $M_{ctr}^k$ represents the mean intra-cluster distance, and $\sigma^k$ represents the standard deviation for the $k^{th}$ emotion category.

\begin{figure}[!tb]
     \centering
    \includegraphics[scale=0.5]{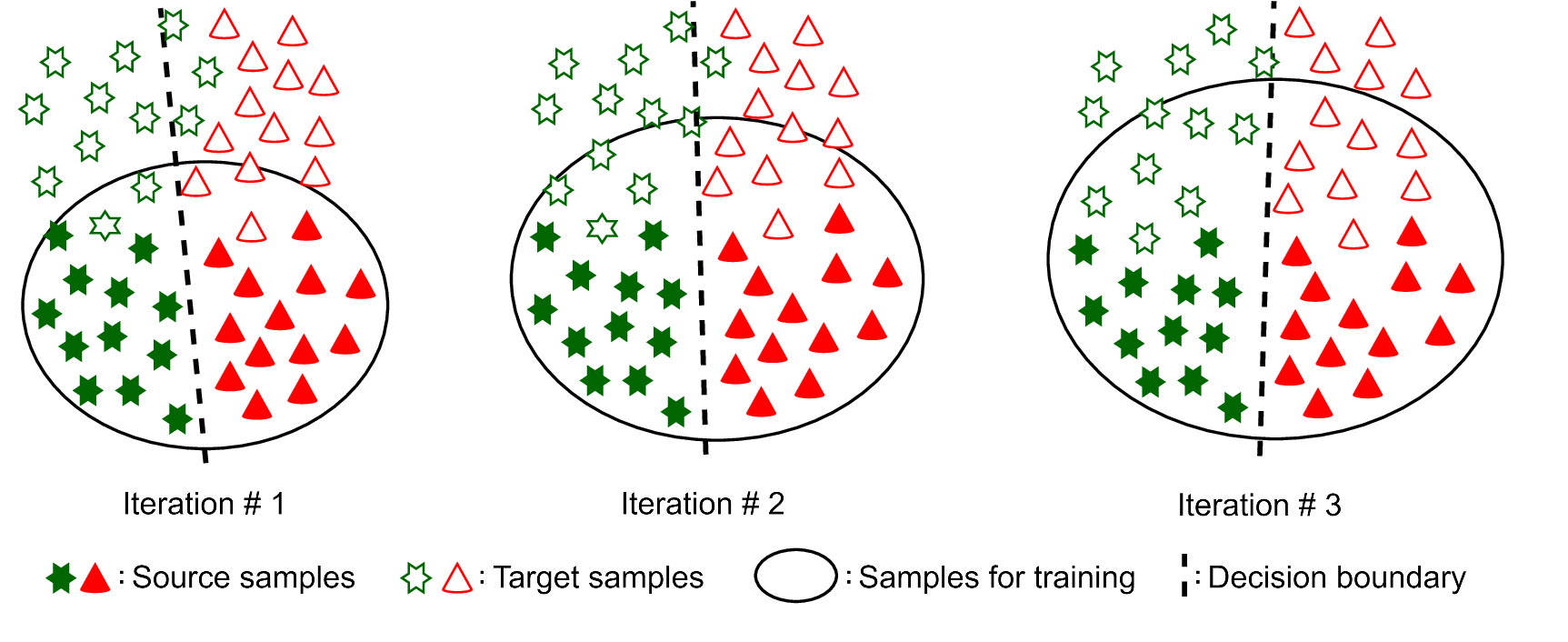}
    \caption{Gradual selection of target domain samples for training.  As the iterations progress and the discrepancy between the source and target domains decreases, target domain samples that were initially excluded (iteration \#1) are considered for training (iterations \#2 and \#3).} 
    \label{fig2}
\end{figure}

Next, from the candidates ($CX_t$), we further refine our selection and choose samples ($SX_t$) with confident predictions, based on softmax values and classifier discrepancies as outlined in the confident prediction conditions in the cluster computation stage. The goal is to select training samples from the target domain that have low distributional differences from the source and are reliable. This is important because using samples that are highly dissimilar from the source or unreliable can lead to negative transfer issues. Subsequently, we perform epoch training using both source domain samples and the selected target domain samples. This process is repeated iteratively until there are no selected samples from the target domain for training or a certain number of interactions have occurred. In each iteration, we check for target samples for training among the excluded samples from the previous iteration. As the iterations progress and the discrepancy between the source and target domains decreases, samples that were initially excluded due to their distance from the source cluster may be selected in later iterations (Figure \ref{fig2}). While GPTDS prioritizes target samples based on their proximity to source clusters and model confidence, it also captures subtle differences between the datasets by gradually broadening the selection as training progresses. This strategy allows the model to incorporate a diverse range of target domain samples, including those that may have initially posed challenges, ultimately leading to a more robust and generalized adaptation.

During epoch training, we minimize the weighted sum of four loss functions: the \textit{cluster-separating loss} ($L_{sep}$), the \textit{cluster-compacting loss} ($L_{comp}$), the \textit{inter-domain cluster discrepancy loss} ($L_{cd}$) (\ref{eq9}), and the \textit{running combined loss} ($L_{cmd}$) (\ref{eq10}), along with the \textit{classification loss} stated in equation (\ref{eq12}), as in our previous model \cite{imtiaz2024cross}.

\begin{align}
L_{cd}= \Sigma_{k=1}^{K}\: D(CC_s^k, CC_t^k)
\label{eq9}
\end{align}

\begin{align}
L_{cmd}= \Sigma_{k=1}^{K}\: D(CC_{m,i}^k, CC_{m}^k) \:\{for \:all \:i : \: 1<= i <= \frac{N_s}{N_b}\}
\label{eq10}
\end{align}

where $N_b$ represents the batch size and 
\begin{align}
CC_{m}^k=avg(CC_s^k, CC_t^k)
\label{eq11}
\end{align}

\begin{align}
\begin{split}
            L =L_{cls} + \beta_1 (L_{comp}^s + L_{comp}^t) + \beta_2 (L_{sep}^s + L_{sep}^t) \\+ \beta_3 L_{cd} + \beta_4 L_{cmb}
\end{split}
\label{eq12}
\end{align}
where $\beta_1$, $\beta_2$, $\beta_3$, and $\beta_4$ denote hyperparameters. 

The \textit{inter-domain cluster discrepancy loss} aims to decrease the distance between clusters in the source and target domains, while the \textit{running combined loss} aims to minimize the distance between the global average cluster centroids (the mean of the cluster centroids from the source and target domains, calculated in the cluster computation stage) and the current average cluster centroids for the current training batch.

\subsubsection{Inference}
During the inference stage, we introduce a Prediction Confidence-aware Test-Time Augmentation (PC-TTA) technique to enhance the model's performance on test data from the target domain. TTA enhances the model's ability to handle unseen variations by applying data augmentation techniques to the test data and combining the predictions. The primary downside of TTA is that applying multiple transformations and performing predictions can be computationally intensive. 
To address this issue, instead of applying TTA to all test samples, we quantify the model's prediction confidence and determine the necessity of applying TTA.

\begin{figure}[!tb]
    \centering
    \includegraphics[scale=0.41]{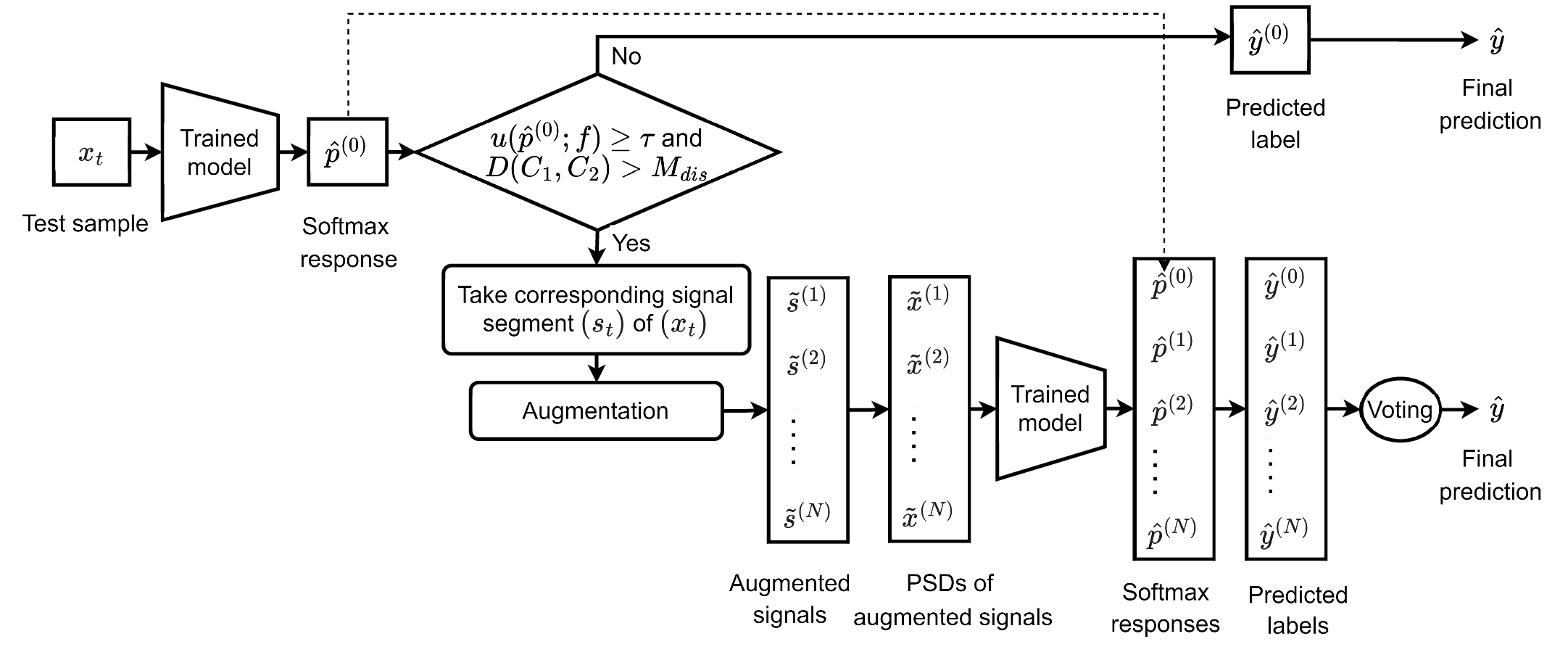}
    \caption{Proposed Prediction Confidence-aware Test-Time Augmentation (PC-TTA) technique.} 
    \label{fig3}
\end{figure}

Figure \ref{fig3} illustrates our PC-TTA process. We obtain the softmax response for each input test sample $x_t \in X_t$ using our trained model $f$. For a test sample $x_t$, the softmax response using $f$ is denoted as $\hat{p}^{(0)} = f(x_t)$. Next, we assess the uncertainty of the prediction by using entropy as the measure, as follows:
\begin{align}
u (\hat{p}^{(0)};f)= - \Sigma_{k=1}^{K}\:  p(y=k|x_t) \log p(y=k|x_t)
\label{eq13}
\end{align}
where $K$ represents the total number of emotion categories, $k$ refers to a specific category within this set, and $p(y=k|x_t)$ denotes the probability of category $k$ for the test sample $x_t$.
 
If $u (\hat{p}^{(0)};f)$ is high, it indicates high uncertainty and reflects the model's poor confidence. When the model's prediction confidence for a test sample is low to some extent, we apply TTA to that sample; otherwise, we accept the model's prediction for that sample as the final output. In our model, when both $u (\hat{p}^{(0)};f)$ and the difference between the two classifiers are high for a test sample, we can infer that the model's confidence in predicting that sample is low. 
If $u (\hat{p}^{(0)};f)$ is greater than or equal to a certain threshold $\tau$, and the classifier discrepancy $D(C_1, C_2)$ is greater than the mean classifier discrepancy $M_{dis}$, then we perform TTA on that sample. Otherwise, we consider the model's prediction as confident, and the predicted label $\hat{y}^{(0)}$ based on the softmax response $\hat{p}^{(0)}$ is considered the final prediction $\hat{y}$.

For the samples that require TTA after this filtering stage, we first extract corresponding EEG signal segment $s_t$ from the input sample $x_t$. Subsequently, we apply augmentation techniques to $s_t$, specifically utilizing Gaussian noise addition and resampling in this experiment. Let the number of transformations be denoted as $N$, resulting in transformed EEG signal segments $\tilde{S} = \{\tilde{s}^{(1)}, \ldots, \tilde{s}^{(N)}\}$. From these signal segments, we extract PSD features $\tilde{X} = \{\tilde{x}^{(1)}, \ldots, \tilde{x}^{(N)}\}$. Next, we feed these augmented features into the pre-trained model $f$ to obtain predicted labels $\{\hat{y}^{(1)}, \ldots, \hat{y}^{(N)}\}$, based on the softmax responses $\{\hat{p}^{(1)}, \ldots, \hat{p}^{(N)}\}$ from $f$. To determine the final prediction, we consider all N+1 predicted labels $\hat{Y} = \{\hat{y}^{(0)}, \hat{y}^{(1)}, \ldots, \hat{y}^{(N)}\}$. Here, $\hat{y}^{(0)}$ corresponds to the softmax response $\hat{p}^{(0)}$ for the original input sample $x_t$. We conduct a vote among these predicted labels, selecting the most frequently occurring class as the final prediction $\hat{y}$.

\begin{algorithm}
\footnotesize
\begin{algorithmic}
\REQUIRE 
\hfill
\\Source PSD samples $X_s$, Target PSD samples $X_t$, Source labels $Y_s$\\
Source signal segments $S_s$, Target  signal segments $S_t$\\
Feature extractor $F$, classifiers $C_1$, $C_2$\\
Epochs $MaxEpoch_1$, $MaxEpoch_2$, $MaxEpoch_3$; $MaxItr$

\ENSURE
\hfill
\\
\textbf{Pre-training} \\
\For{($i$ = 1 to $MaxEpoch_1$)}{
    Train the model with $X_s$ and $Y_s$:\\
    Update the parameters of $F$, $C_1$, and $C_2$ by minimizing L (\ref{eq1})
}
\textbf{Cluster Computation}\\
Calculate source cluster centroids $CC_{s_{pre}}$\\
\For{($i$ = 1 to $MaxEpoch_2$)}{
    Reduce intra-cluster distance and increase inter-cluster distance:\\
    Update the parameters of $F$, $C_1$, and $C_2$ by minimizing L (\ref{eq4})
}
Calculate source cluster centroids $CC_s$\\
Calculate mean intra-cluster distance $M_{ctr}$, standard deviation $\sigma$, and \\mean classifier discrepancy $M_{dis}$ using Eq. (\ref{eq5},\ref{eq6}, \ref{eq7})\\
Calculate target cluster centroids $CC_t$\\

\textbf{Domain Adaptation}\\
Select candidates $CX_t$ for training from $X_t$ using Eq. (\ref{eq8})\\
Select samples $SX_t$ for training from $CX_t$ with confident predictions\\
\While{($SX_t$ is not empty and iteration $<=$ MaxItr)}{

    $X_t$ = \{$X_t$ - $SX_t$\}\\
    \For{($i$ = 1 to $MaxEpoch_3$)}{
        Train the model with $SX_t$, $X_s$ and $Y_s$:\\
        Update the parameters of $F$, $C_1$, and $C_2$ by minimizing L (\ref{eq12})
        }
    Select candidates $CX_t$ for training from $X_t$ using Eq. (\ref{eq8})\\
    Select samples $SX_t$ for training from $CX_t$ with confident predictions\\
}
\textbf{Inference}\\
\For{(each $x_t$ $\in$ $X_t$)}{
    \If{(model's prediction uncertainty $u (\hat{p}^{(0)};f)$ $>=$ threshold $\tau$ and \\ \: \:  classifier discrepancy $D(C_1, C_2)$ $>$ mean classifier discrepancy $M_{dis}$)}{

            Extract signal segment ($s_t$) corresponding to ($x_t$)\\
            Augment ($s_t$) and create $N$ transformed EEG segments $\tilde{S}$\\
            Extract $N$ PSD features $\tilde{X}$ from $\tilde{S}$\\
            Obtain $N+1$ labels $\hat{Y}$ predicted by the model for both $\tilde{X}$ and $x_t$\\
            Perform majority voting on $N+1$ predictions to obtain final prediction $\hat{y}$          
    }
    \Else{
    Consider the label of $x_t$ predicted by the model as final prediction $\hat{y}$.
    }

}
\caption{Steps of the proposed unsupervised domain adaptation method for cross-dataset EEG-based emotion recognition.}
\label{alg}
\end{algorithmic}
\end{algorithm}

Algorithm \ref{alg} outlines the complete procedure of our proposed approach.

\subsection{Data preprocessing and construction of model inputs}

Power Spectral Density (PSD) \cite{li2023tmlp+,rayatdoost2021subject,li2019domain} and Differential Entropy (DE) \cite{jimenez2023learning,she2023cross,shen2022contrastive} features are widely employed in EEG-based emotion recognition and have demonstrated superior performance compared to other EEG features in previous studies \cite{ni2021domain,li2022eeg}. In this study, we explore both PSD and DE; however, our experimental results indicate that PSD outperforms DE. As a result, we incorporate PSD into our proposed method. The formula for computing the PSD is as follows:

\begin{align}
PSD(f)= \frac{1}{N} \Big| \Sigma_{n=1}^{N} \: x(n) \: \exp ^{-j2\pi fn}\Big| ^2
\label{eq14}
\end{align}
where, $PSD(f)$ represents the PSD at frequency $f$, $x(n)$ represents the signal segment, and $N$ corresponds to the number of samples in $x(n)$.
\begin{align}
PSD_{band}= \int_{f_{low}}^{f_{high}}PSD(f) \, df
\label{eq15}
\end{align}
where, $PSD_{band}$ represents the PSD in the frequency band [$f_{low}$, $f_{high}$].

\begin{figure}[!tb]
     \centering
    \includegraphics[scale=0.5]{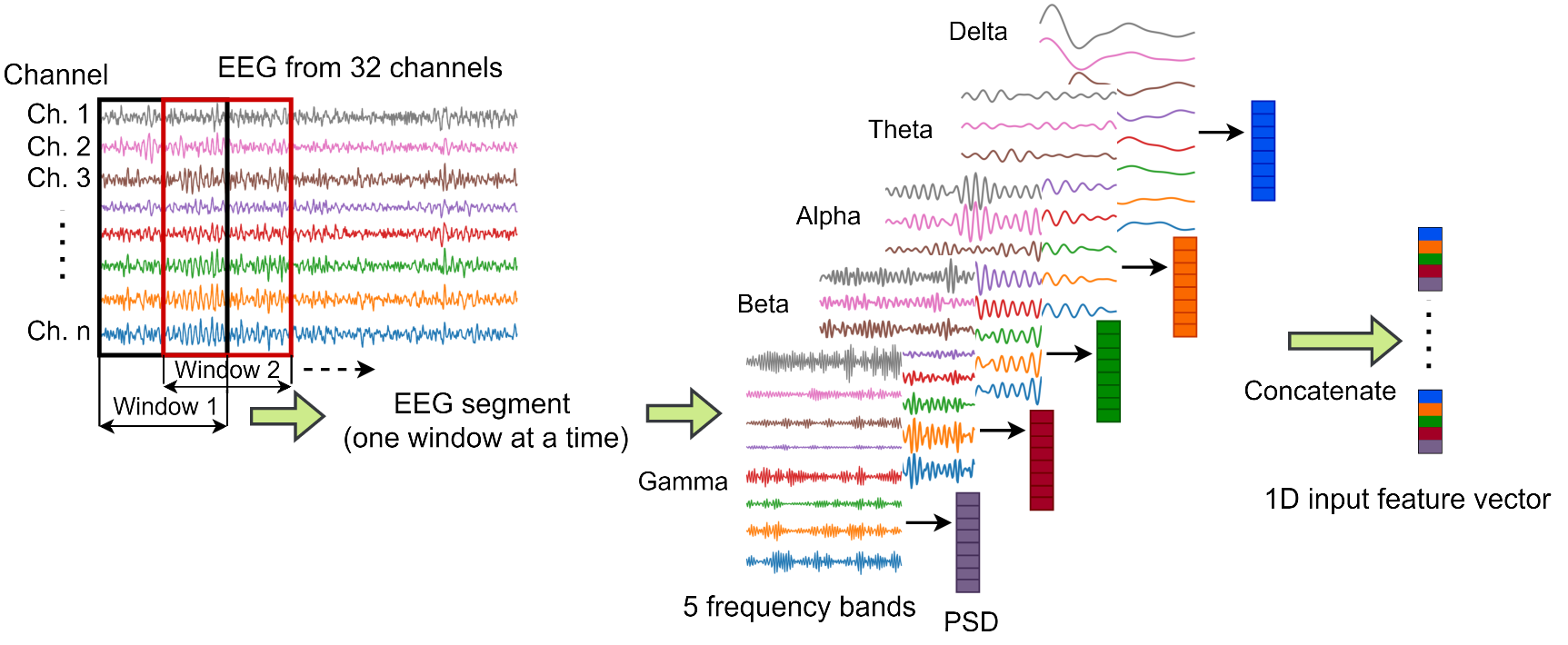}
    \caption{Genaration of 1D input features.} 
    \label{fig4}
\end{figure}

Figure \ref{fig4} illustrates the process of constructing the temporal input for our model. To ensure a uniform input size for the model, we extract EEG signals from 32 common EEG channels present in both the DEAP and SEED datasets. Before feature extraction, we divide each EEG trial into multiple segments. For segmentation, we use a window size of 2 seconds and a step size of 1 second, similar to \cite{wang2021prototype}, resulting in a 1-second overlap between segments. Next, we extract the PSD features from each segment in each channel at frequency bands: delta $\delta$ (1–3 Hz), theta $\theta$ (4–7 Hz), alpha $\alpha$ (8–13 Hz), beta $\beta$ (14–30 Hz), and gamma $\gamma$ (31–50 Hz).  We then create a 1-dimensional feature vector $X \in \mathbb{R}^{n*i}$ by concatenating the PSD features. Here, $n$=32 represents the number of channels, and $i$ is set to 5, corresponding to the frequency bands $\delta$, $\theta$, $\alpha$, $\beta$, and $\gamma$. Therefore, the size of the 1-dimensional input feature vector is $5 \times 32 = 160$. Finally, we normalize the features to fall within the range of [-1, 1]. To address potential discrepancies between feature spaces, we ensure uniform preprocessing and normalization across domains, aligning data transformations and reducing the risk of errors.

\section{Experiments}
\label{sec:experiments}

\subsection{Datasets and experimental setup}
Our proposed model has been evaluated through experiments using the DEAP and SEED datasets, which are widely utilized for emotion recognition tasks and are publicly available. Our experiments include cross-dataset testing, where the model is trained on one dataset and evaluated on another.

\subsubsection{DEAP dataset \cite{koelstra2011deap}}
The stimuli consisted of 40 one-minute music videos. The experiment included 32 healthy participants. Each subject underwent 40 trials, each lasting 63 seconds. This included a 3-second pre-trial period and 60 seconds of watching one-minute videos. After viewing each video, participants rated their arousal, valence, dominance, and liking using a continuous scale ranging from 1 to 9. EEG signals were recorded using 32 electrodes at a sampling frequency of 512 Hz. The signals were downsampled to 128 Hz during preprocessing and further filtered using a band-pass filter between 4 Hz and 45 Hz to reduce noise and artifacts. 

\subsubsection{SEED dataset \cite{zheng2015investigating}}
A total of 15 Chinese movie clips, each carefully selected and approximately 4 minutes long, were used as stimuli. Fifteen Chinese subjects participated in the study, comprising 7 females and 8 males. Each participant attended 3 sessions on different days, during which they watched the same set of 15 movie clips. The movie clips were intended to elicit three emotions: positive, neutral, and negative. EEG signals were captured using 62-channel electrode caps at a sampling frequency of 1000 Hz. The recorded data underwent preprocessing, including downsampling to 200 Hz and filtering with a bandpass of 0-75 Hz.
\\

In our experiments, we utilize data samples from all subjects and trials in both the SEED and DEAP datasets. Since the first 3 seconds of each trial in the DEAP dataset consist of pre-trial period data, we exclude this initial 3-second segment from each trial. We filter the signals from the SEED dataset using a bandpass filter with a passband of 0.3 Hz to 50 Hz to remove noise and artifacts, following \cite{li2019domain,li2022dynamic}. Since the DEAP dataset signals are already filtered between 4 Hz and 45 Hz, we do not perform additional filtering. This study focuses on two emotion categories: positive and negative. In the DEAP dataset, valence values above 4.5 are considered positive, and those below 4.5 are considered negative based on manual classification guidelines \cite{ni2021domain,salankar2021emotion}. For the SEED dataset, we exclude neutral-labeled samples and retain only those labeled as positive or negative for analysis. After preprocessing, the DEAP dataset contains 74,240 samples, while the SEED dataset contains 99,630 samples, each with a vector size of 160. All samples from the source domain dataset are used for training, and all samples from the target domain dataset are used for testing.

All experiments are conducted on a Linux platform using Python (version 3.10.12) and the PyTorch library (version 2.3.1+cu121), running on an NVIDIA Tesla T4 GPU with 12GB of memory. The learning rate is set to 0.001, and weight decay is set to 0.0005, following our previous work \cite{imtiaz2024cross}. The batch size is set to 64, the Rectified Linear Unit (ReLU) is used as the activation function, and model optimization is performed using the Adam optimizer. The hyperparameters $\alpha$, $\gamma_1$, $\gamma_2$, $\beta_1$, $\beta_2$, $\beta_3$, and $\beta_4$ are set to 0.5, 0.1, 0.1, 0.1, 0.1, 0.5, and 0.1, respectively, consistent with our previous method \cite{imtiaz2024cross}.

\subsection{Results and discussion}
Our proposed method is evaluated in a cross-dataset setting, where the source domain (SD) and target domain (TD) are from different datasets. We test our model by training on one dataset (SD) and testing on the other (TD), in both directions. Specifically, we train on the DEAP dataset and test on the SEED dataset, and vice versa. For comparison with existing approaches, we shortlisted the existing works according to two categories: 1) existing domain adaptive emotion recognition methods that provide experimental results on the same datasets with identical training and testing settings, and 2) existing state-of-the-art general domain adaptive approaches that are open-source, so that we can easily re-implement for comparison.  

For category 1, to the best of our knowledge, only two prior studies on EEG-based emotion recognition\cite{gu2022multi,ni2021domain} have used the same training and testing settings as ours. Table \ref{table2} presents a comparison between our proposed method and the prior approaches, demonstrating our method's superior performance by a significant margin. Specifically, our method achieves 59.68\% accuracy when tested on DEAP (trained on SEED), outperforming the previous best of 53.67\% by Gu et al. \cite{gu2022multi}. Similarly, our method achieves 67.44\% accuracy when tested on SEED (trained on DEAP), surpassing the previous best of 64.97\% by Ni et al. \cite{ni2021domain}.

For category 2 (open source projects), we compare our proposed approach against six state-of-the-art domain-adaptive approaches \cite{chen2021ms,li2019domain,sagawa2019distributionally,huang2020self,zhi2024confusing,jimenez2023learning}. To ensure a fair comparison, we re-implemented these approaches using their open-source repositories and adopted identical network architecture and experimental setup as our proposed approach. Table \ref{table2} demonstrates the superior performance of our proposed method compared to the domain-adaptive approaches in terms of overall accuracy. Our method outperforms the second-best accuracy achieved by Jiménez-Guarneros and Fuentes-Pineda \cite{jimenez2023learning} by 6.20\% for SEED $\rightarrow$ DEAP (trained on SEED, tested on DEAP). Similarly, our method surpasses the second-best accuracy achieved by Chen et al. \cite{chen2021ms} by 2.72\% for DEAP $\rightarrow$ SEED. Furthermore, we conduct a paired-sample \textit{t-test}, using \textit{p-values}, to determine the significance of the differences in emotion recognition performance between our proposed method and other approaches. The results demonstrate that the difference in accuracy between our method and all other approaches is highly significant (**) for SEED $\rightarrow$ DEAP. For DEAP $\rightarrow$ SEED, the difference is significant (*) compared to Chen et al. \cite{chen2021ms}, Li et al. \cite{li2019domain}, and Jiménez-Guarneros and Fuentes-Pineda \cite{jimenez2023learning}, while the difference is highly significant compared to other approaches.

\begin{table*} [!tb]
\scriptsize
\centering
\caption {Comparison of overall accuracy values between our proposed method and other state-of-the-art domain-adaptive methods. The results for the top two methods \cite{ni2021domain,gu2022multi} are directly drawn from the papers due to identical training-test settings, while the remaining methods are implemented and evaluated using the same network architecture and experimental setup as ours. Symbols indicate differences in accuracies (paired-sample \textit{t-test}: $\sim$ nonsignificant, *p $<$ 0.05, **p $<$ 0.01).}
\begin{tabular}
{ p{0.43\linewidth} p{0.24\linewidth} p{0.21\linewidth}}
\hline
  & \multicolumn{2}{c}{Accuracy (\%)}\\
\hline

 & SEED\cite{zheng2015investigating}$\rightarrow$DEAP\cite{koelstra2011deap} & DEAP\cite{koelstra2011deap}$\rightarrow$SEED\cite{zheng2015investigating}  \\

\hline
Ni et al. \cite{ni2021domain}  & 53.54 & 64.97 \\
Gu et al. \cite{gu2022multi} & 53.67 & 64.67 \\
Li et al. \cite{li2019domain}  & 53.44** & 63.88* \\
Chen et al. \cite{chen2021ms}  & 51.93** & 64.72* \\
Sagawa et al. \cite{sagawa2019distributionally} & 48.91** & 58.39**\\
Huang et al. \cite{huang2020self}  & 52.56** &  59.20**\\
Jiménez-Guarneros and Fuentes-Pineda  \cite{jimenez2023learning}   & 53.48** & 63.45*\\
Zhi et al. \cite{zhi2024confusing}  & 51.89** &  61.56**\\
\textbf{Proposed method} & \textbf{59.68} & \textbf{67.44} \\
\hline
\end{tabular}
\label{table2}
\end{table*}

\begin{figure*}[!htb]
     \centering
    \begin{tabular}{cc}
\subfloat[]{\includegraphics[scale=0.3]{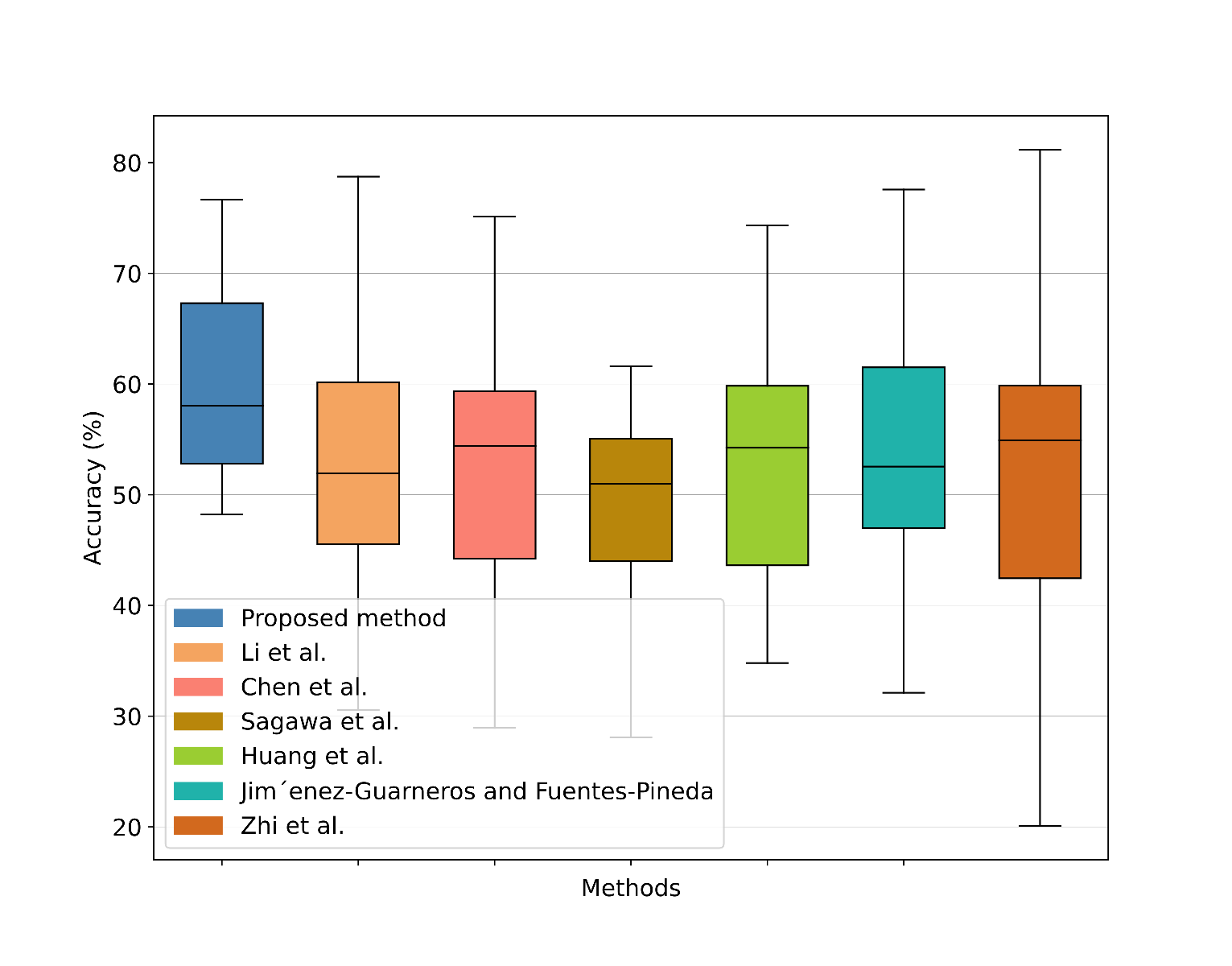}}
\subfloat[]{\includegraphics[scale=0.3]{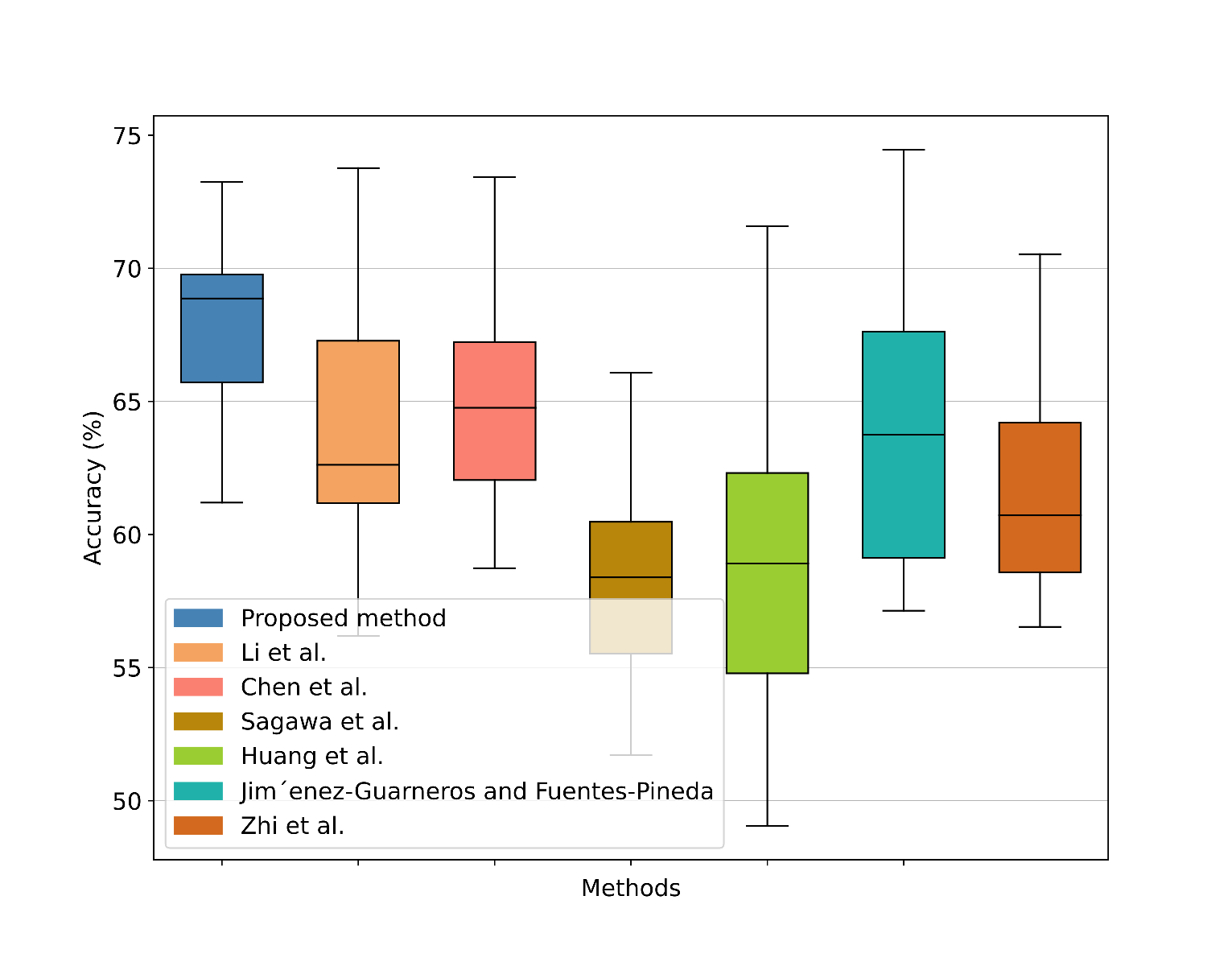}}  
\end{tabular}
     
        \caption{Boxplots showing the distribution of emotion recognition accuracies for our proposed method and other domain-adaptive methods. (a) SEED\cite{zheng2015investigating} $\rightarrow$ DEAP\cite{koelstra2011deap} (b) DEAP\cite{koelstra2011deap} $\rightarrow$ SEED\cite{zheng2015investigating}.}
        \label{fig5}
\end{figure*}

Figure \ref{fig5} displays the accuracy distributions of our proposed method and six domain-adaptive methods using boxplots. Our method stands out by achieving the highest median accuracy among all methods. When trained on SEED and tested on DEAP, Sagawa et al.'s method exhibits relatively less variation, but its overall accuracy is low. In contrast, our method achieves the highest median accuracy of 58.03\%, with the majority of prediction accuracies exceeding this value. In the DEAP $\rightarrow$ SEED scenario, our method surpasses all others, achieving the highest median accuracy (68.87\%) with low variations in accuracies compared to alternative methods. 

\begin{table*} [!tb]
\scriptsize
\centering
\caption {Performance comparison of our proposed method and other approaches for identifying positive and negative emotions.}
\begin{tabular}
{p{0.24\linewidth} p{0.04\linewidth} p{0.04\linewidth}  p{0.04\linewidth} p{0.04\linewidth} p{0.04\linewidth} p{0.04\linewidth} p{0.04\linewidth} p{0.04\linewidth} p{0.04\linewidth} p{0.04\linewidth} p{0.04\linewidth} p{0.045\linewidth}  }
\hline
 &  \multicolumn{6}{c}{SEED\cite{zheng2015investigating}$\rightarrow$DEAP\cite{koelstra2011deap}}  & \multicolumn{6}{c}{DEAP\cite{koelstra2011deap}$\rightarrow$SEED\cite{zheng2015investigating}}\\
\cline{2-13}
\hline
 &  & Negative & & & Positive & & & Negative & & & Positive & \\
\cline{2-13}
& Se (\%)& PPV (\%)& F1 (\%) & Se (\%) & PPV (\%) & F1 (\%) & Se (\%) & PPV (\%) & F1 (\%) & Se (\%) & PPV (\%) & F1 (\%)\\
\hline
Li et al. \cite{li2019domain} & 27.52 & 33.87 & 30.37 & 68.60 & 61.82 & 65.03 & 29.79 & 93.56 & 45.19 & 97.95 & 58.26 & \textbf{73.06}\\

Chen et al. \cite{chen2021ms} & 29.97 & 33.21 & 31.51& 64.77 & 61.27 & 62.97 & 46.17 & 73.38 & 56.68 & 83.26 & 60.75 & 70.25\\

Sagawa et al. \cite{sagawa2019distributionally} & 37.22 & 32.96 & 34.96 & 55.74 & 60.30 & 57.93 & 61.72 & 57.85 & 59.72 & 55.06 & 59.01 & 56.97\\

Huang et al. \cite{huang2020self}  & 30.84 & 34.16 & 32.42 & 65.25 & 61.74 & 63.45 & 47.35 & 62.04 & 53.71 & 71.05 & 57.45 & 63.53\\

Jiménez-Guarneros and Fuentes-Pineda  \cite{jimenez2023learning} & 33.92 & 36.11 & 34.98 & 64.91 & 62.69 & 63.78 & 43.72 & 72.20 & 54.46 & 83.18 & 59.66 & 69.48\\

Zhi et al. \cite{zhi2024confusing} & 28.86 & 32.75 & 30.68 & 65.35 & 61.11 & 63.16 & 38.93 & 71.09 & 50.31 & 84.18 & 57.97 & 68.66\\

Imtiaz and Khan \cite{imtiaz2024cross} & 26.29 & 33.57 & 29.49 & 69.58 & 61.76 & 65.44 & 46.45 &  64.31 & 53.94 & 74.24 & 58.11 & 65.19\\

\textbf{Proposed method} & 29.70 & 43.24 & \textbf{35.21} & 77.21 & 65.26 & \textbf{70.73} & 50.85 & 76.08 & \textbf{60.96} & 84.02 & 63.11 & 72.08\\
\hline
\end{tabular}
\label{table3}
\end{table*}

Table \ref{table3} compares the sensitivity, positive predictive value (PPV), and F1 score of our proposed method with six domain-adaptive approaches and our previous model. While Li et al.'s method yields the highest F1 score of 73.06\% (0.98\% higher than ours) in recognizing positive emotions, our proposed approach excels in recognizing negative emotions when tested on SEED. In contrast, when tested on DEAP, all methods face challenges in recognizing negative emotions. Nonetheless, our proposed method outperforms other methods, including our previous model, in recognizing both negative and positive emotions. The lower performance in recognizing negative emotions on DEAP may be attributed to the smaller number of samples with negative emotion (36.87\%) compared to those with positive emotion (63.13\%). In contrast, the SEED dataset features a balanced distribution of samples (negative- 50.36\%, positive- 49.64\%).

\begin{table*} [!tb]
\scriptsize
\centering
\caption {Comparison of model sizes between our proposed method and other approaches (approximate values indicated by *).}
\begin{tabular}
{ p{0.43\linewidth} p{0.23\linewidth} }
\hline
Model & Number of Parameters\\
\hline
Li et al. \cite{li2019domain} & 40,453 \\
Chen et al. \cite{chen2021ms} & 122,963 \\
Sagawa et al. \cite{sagawa2019distributionally} & 23,555,098* \\
Huang et al. \cite{huang2020self} & 34,583,605* \\
Jiménez-Guarneros and Fuentes-Pineda  \cite{jimenez2023learning} & 1,088,387 \\
Zhi et al. \cite{zhi2024confusing} & 23,792,099* \\
Imtiaz and Khan \cite{imtiaz2024cross} & 357,392 \\
\textbf{Proposed method} & \textbf{34,288} \\
\hline
\end{tabular}
\label{table4}
\end{table*}

Table \ref{table4} compares model sizes, measured by the number of parameters, between our proposed method and other approaches. We report average model sizes for the other methods, as they employed different architectures for various datasets. The model sizes for Ni et al. \cite{ni2021domain} and Gu et al. \cite{gu2022multi} are not included due to the unavailability of this information. Our model is the lightest among those compared, significantly lighter than the complex, heavyweight models used by others, while still achieving superior performance. This reduction in model size leads to lower computational complexity in terms of both time and memory. Specifically, our efficient architecture reduces the number of parameters to 34,288, enabling faster training and inference times with lower memory usage.

\subsubsection{Ablation study}

\begin{figure*}[p]
     \centering
    \begin{tabular}{cc}
\subfloat[]{\includegraphics[scale=0.41]{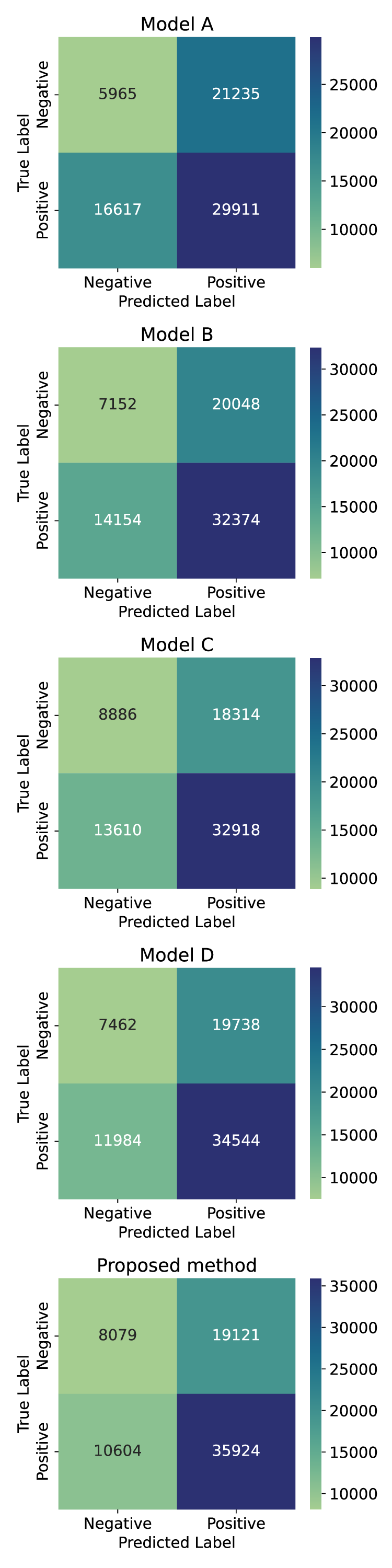}} \; \;
\subfloat[]{\includegraphics[scale=0.41]{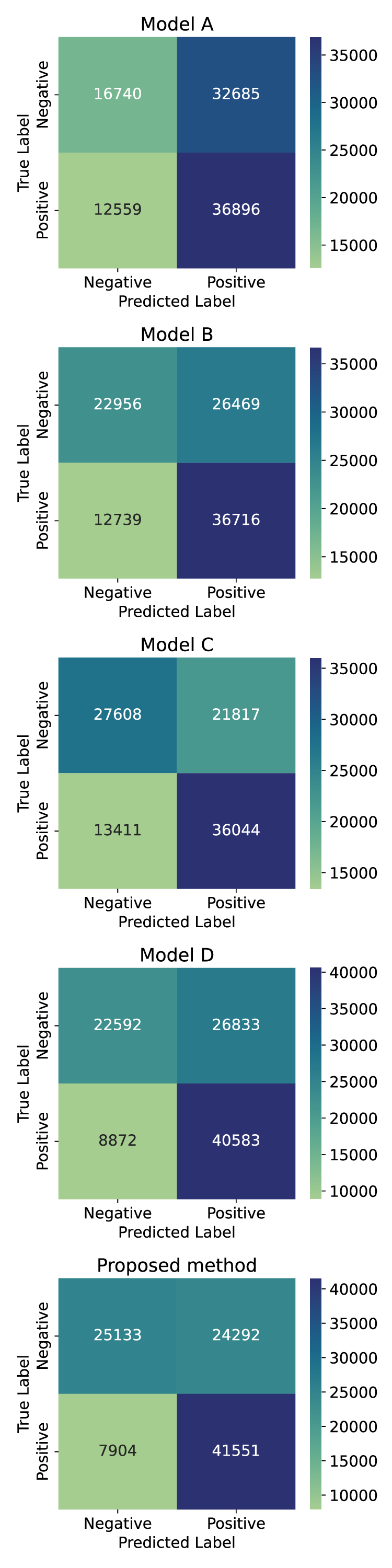}}  
\end{tabular}
     
        \caption{Confusion matrices from the ablation study for (a) SEED\cite{zheng2015investigating} $\rightarrow$ DEAP\cite{koelstra2011deap} and (b) DEAP\cite{koelstra2011deap} $\rightarrow$ SEED\cite{zheng2015investigating}.}
        \label{fig6}
\end{figure*}

We perform an ablation study to determine how each component of our proposed method affects the results. We systematically remove the main components of our proposed method, one at a time, which constitute the primary contributions of this research. We then assess the model and observe the effect of removing each component. Table \ref{table5}, Table \ref{table6}, and Figure \ref{fig6} present the results of the ablation analysis. We create four models by excluding one component at a time while leaving all other components unchanged. Model A is created by removing all domain adaptation components from the proposed model, retaining only the pre-training stage. Model B represents our previously proposed method \cite{imtiaz2024cross}, referred to as the \textit{baseline model}. Model C excludes the Gradual Proximity-guided Target Data Selection (GPTDS) component from the proposed method. Model D is constructed by excluding the Prediction Confidence-aware Test-Time Augmentation (PC-TTA) component while keeping all other components unchanged. 

In addition to evaluating the models using Power Spectral Density (PSD), we also assess them using Differential Entropy (DE) features. Table \ref{table5} displays the accuracy of all models for both PSD and DE features. Each component influences performance to some extent. Overall, we achieve better performance when using PSD features, although DE works slightly better for Model C in SEED $\rightarrow$ DEAP and for Model B in DEAP $\rightarrow$ SEED. In most cases, PSD performs well. Therefore, we choose PSD for our proposed approach and focus on comparing others using PSD from here on. Model A performs the worst, even lower than the random probability for SEED $\rightarrow$ DEAP, as it lacks any knowledge about the test (target domain) data. The \textit{baseline model} achieves an accuracy of 53.61\% for SEED $\rightarrow$ DEAP and 60.35\% for DEAP $\rightarrow$ SEED. Both GPTDS and PC-TTA significantly impact the performance of emotion classification. PC-TTA improves the accuracy compared to the \textit{baseline model} by 3.09\% for SEED $\rightarrow$ DEAP and by 4.02\% for DEAP $\rightarrow$ SEED. GPTDS improves the accuracy compared to the \textit{baseline model} by 3.36\% for SEED $\rightarrow$ DEAP and by 3.54\% for DEAP $\rightarrow$ SEED.

\begin{table*} [!tb]
\scriptsize
\centering
\caption {Overall accuracy comparisons on both datasets through the ablation study.}
\begin{tabular}
{p{0.24\linewidth} p{0.16\linewidth} p{0.16\linewidth}  p{0.16\linewidth} p{0.06\linewidth} }
\hline
&  \multicolumn{4}{c}{Accuracy (\%)} \\
\hline
 & SEED\cite{zheng2015investigating}$\rightarrow$DEAP\cite{koelstra2011deap} &  & DEAP\cite{koelstra2011deap}$\rightarrow$SEED\cite{zheng2015investigating} &\\
\cline{2-5}
& PSD & DE & PSD  & DE \\
\hline
Model A & 48.66 & 48.03 & 54.24 & 54.07\\

Model B & 53.61 & 53.53 & 60.35 & 60.58 \\

Model C & 56.70 & 56.89 & 64.37 & 63.90 \\

Model D & 56.97 & 55.86 & 63.89 & 61.75 \\

\textbf{Proposed method} & \textbf{59.68} & 59.02 & \textbf{67.44} & 66.26 \\
\hline
\end{tabular}
\label{table5}\end{table*}

\begin{table*} [!tb]
\scriptsize
\centering
\caption {Performance of different components in the proposed method for identifying positive and negative emotions.}
\begin{tabular}
{p{0.21\linewidth} p{0.04\linewidth} p{0.04\linewidth}  p{0.04\linewidth} p{0.04\linewidth} p{0.04\linewidth} p{0.04\linewidth} p{0.04\linewidth} p{0.04\linewidth} p{0.04\linewidth} p{0.04\linewidth} p{0.04\linewidth} p{0.05\linewidth}  }
\hline
 &  \multicolumn{6}{c}{SEED\cite{zheng2015investigating}$\rightarrow$DEAP\cite{koelstra2011deap}}  & \multicolumn{6}{c}{DEAP\cite{koelstra2011deap}$\rightarrow$SEED\cite{zheng2015investigating}} \\
\cline{2-13}
\hline
 &  & Negative & & & Positive & & & Negative & & & Positive & \\
\cline{2-13}
& Se (\%)& PPV (\%)& F1 (\%) & Se (\%) & PPV (\%) & F1 (\%) & Se (\%) & PPV (\%) & F1 (\%) & Se (\%) & PPV (\%) & F1 (\%)\\
\hline
Model A & 21.93 & 26.41 & 23.96 & 64.29 & 58.48 & 61.25 & 33.87 &  57.14 &  42.53 & 74.61 & 53.03 & 62.00\\

Model B & 26.29 & 33.57 & 29.49 & 69.58 & 61.76 & 65.44 & 46.45 &  64.31 & 53.94 & 74.24 & 58.11 & 65.19\\

Model C & 32.67 & 39.50 & \textbf{35.76} & 70.75 & 64.25 & 67.34 & 55.86 & 67.31 & \textbf{61.05} & 72.88 & 62.29 & 67.17\\

Model D & 27.43 & 38.37 & 31.99 & 74.24 & 63.64 & 68.53 & 45.71 & 71.80 & 55.86 & 82.06 & 60.20 & 69.45\\

\textbf{Proposed method} & 29.70 & 43.24 & 35.21 & 77.21 & 65.26 & \textbf{70.73} & 50.85 & 76.08 & 60.96 & 84.02 & 63.11 & \textbf{72.08}\\
\hline
\end{tabular}
\label{table6}
\end{table*}

GPTDS and PC-TTA components enhance the model's performance in classifying both positive and negative emotions, as demonstrated in Table \ref{table6} and Figure \ref{fig6}. The confusion matrices (Figure \ref{fig6}) show that PC-TTA significantly improves the identification of negative emotions for both datasets, while GPTDS significantly enhances the identification of positive emotions for both datasets compared to the \textit{baseline model}. Table \ref{table6} indicates that for classifying negative emotions, PC-TTA achieves the highest F1 score, surpassing even the proposed method, with scores of 35.76\% and 61.05\% for SEED $\rightarrow$ DEAP and DEAP $\rightarrow$ SEED, respectively. The combined use of GPTDS and PC-TTA substantially boosts performance over the \textit{baseline model} for identifying both positive and negative emotions.

\subsubsection{Analysis of PC-TTA threshold ($\tau$)}

\begin{figure*}[!htb]
     \centering
    \begin{tabular}{cc}
\subfloat[]{\includegraphics[scale=0.4]{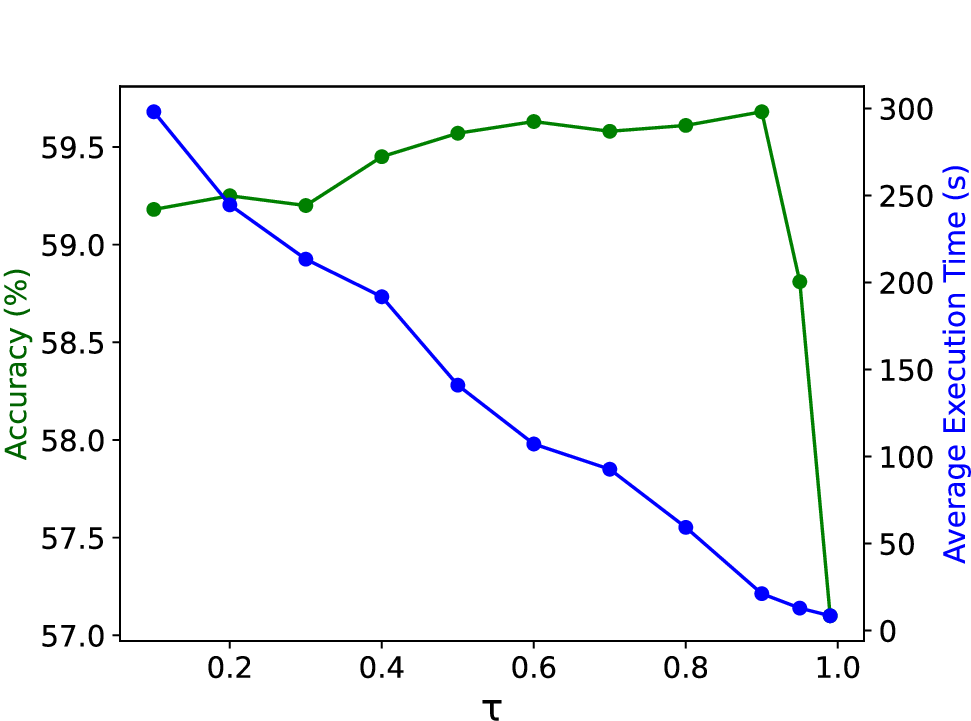}} \; \;
\subfloat[]{\includegraphics[scale=0.4]{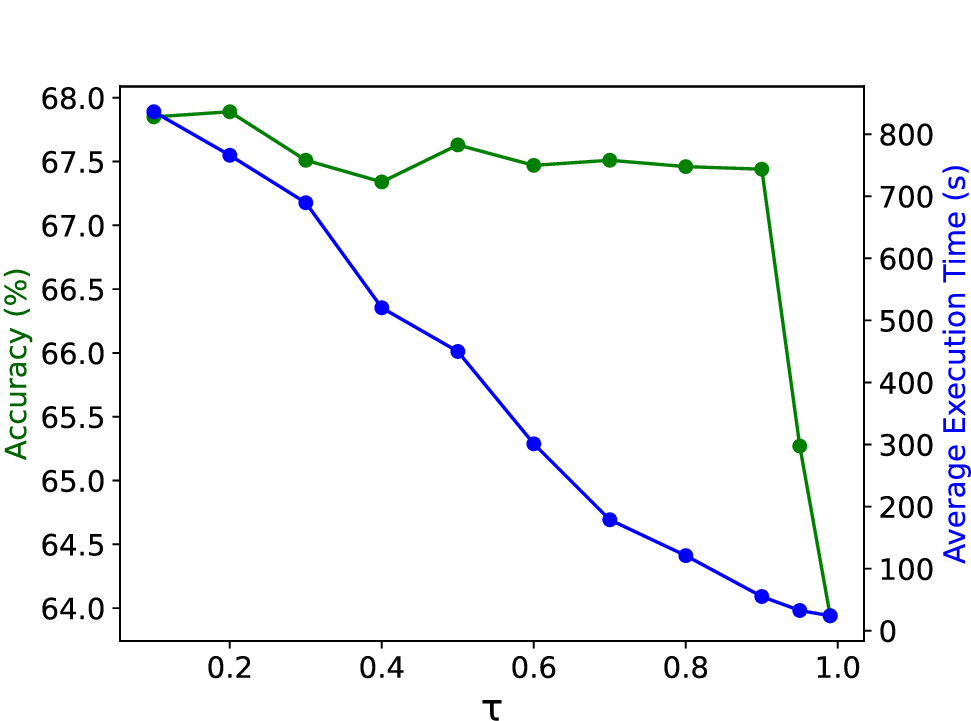}}  
\end{tabular}
     
        \caption{Model performance for different values of $\tau$. (a) SEED\cite{zheng2015investigating} $\rightarrow$ DEAP\cite{koelstra2011deap} (b) DEAP\cite{koelstra2011deap} $\rightarrow$ SEED\cite{zheng2015investigating}.}
        \label{fig7}
\end{figure*}

To select the optimal value of threshold $\tau$, which is used to determine whether to perform TTA, we evaluate the model by varying its value from 0.1 to 0.99. Figure \ref{fig7} illustrates the overall accuracy and average execution time per subject during the inference stage for different values of $\tau$. After careful consideration of both emotion recognition accuracy and computational cost, we have selected $\tau = 0.9$ for our proposed method. The model achieves its highest accuracy when tested on the DEAP dataset with $\tau = 0.9$. However, when tested on SEED, although the model achieves slightly better accuracy with lower $\tau$, the associated high computational time becomes a significant concern.

Table \ref{table7} compares the performance of our PC-TTA approach with two other scenarios: full TTA, where TTA is applied to all test samples, and no TTA. In both cases, applying TTA (PC-TTA and full TTA) significantly improves the emotion recognition accuracy. However, it is crucial to consider the significant computational cost associated with TTA. This cost arises from a series of operations executed each time TTA is applied to a test sample, including extracting the corresponding signal segment, performing a number of augmentations, extracting PSD features from the transformed signal segments, and obtaining the model prediction by feeding them into the model. The execution time of our proposed PC-TTA approach is 2.86 and 2.43 times higher than when no TTA is applied, for the DEAP and SEED test datasets, respectively. This increase is significantly greater when applying TTA to all test samples, with execution times 43.77 and 37.77 times higher when tested on DEAP and SEED, respectively. While full TTA achieves slightly better accuracy (0.34\% higher than our proposed PC-TTA) when tested on SEED, our PC-TTA performs better than full TTA when tested on DEAP. Therefore, our PC-TTA approach effectively balances computational cost while maintaining high classification accuracy.

\begin{table*} [!htb]
\scriptsize
\centering
\caption {Comparison of PC-TTA (proposed), Full TTA, and No TTA. }
\begin{tabular}
{p{0.14\linewidth} p{0.13\linewidth} p{0.20\linewidth}  p{0.13\linewidth} p{0.19\linewidth} }
\hline
 &   \multicolumn{2}{c}{SEED\cite{zheng2015investigating}$\rightarrow$DEAP\cite{koelstra2011deap}} & \multicolumn{2}{c}{DEAP\cite{koelstra2011deap}$\rightarrow$SEED\cite{zheng2015investigating}}\\
\cline{2-5}
& Accuracy (\%) & Average Execution Time (s) & Accuracy (\%) & Average Execution Time (s)\\
\hline

Full TTA & 59.16 & 323.9 & 67.78 & 857.4\\

No TTA & 56.97 & 7.4 & 63.89 & 22.7 \\

\textbf{PC-TTA (Proposed)} & 59.68 & 21.2 & 67.44 & 55.1\\ 

\hline
\end{tabular}
\label{table7}
\end{table*}

\subsubsection{Frequency band analysis in emotion recognition}

\begin{figure}[!htbp]
    
    \includegraphics[scale=0.7]{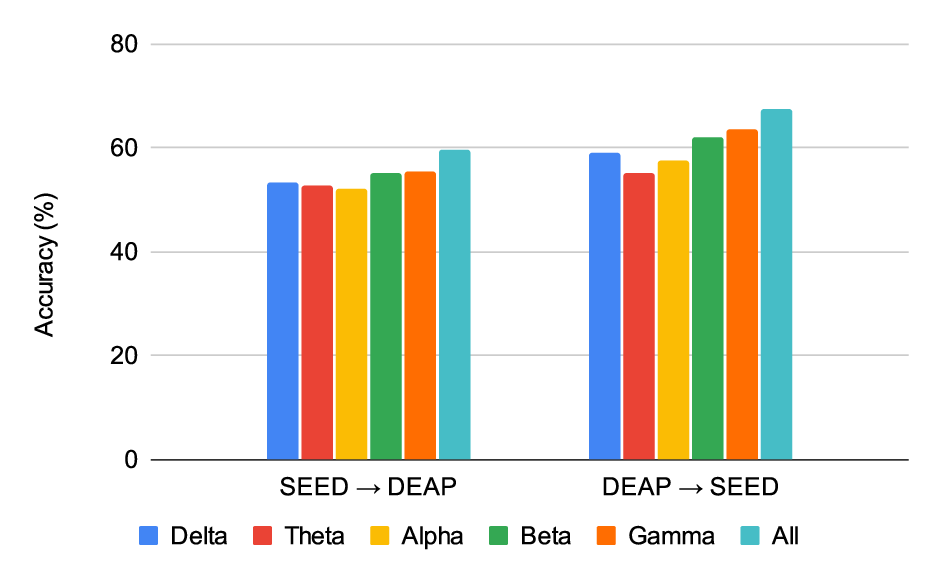}
    \caption{Contribution of frequency bands to emotion recognition.} 
    \label{fig8}
\end{figure}

To assess the impact of each frequency band on emotion recognition, we conduct experiments for each individual frequency band. We adjusted the input layer of the feature extractor ($F$) to accommodate the change in input features from 5 bands to 1 band. Notably, the gamma and beta bands contribute more significantly to emotion recognition compared to the other frequency bands, as illustrated in Figure \ref{fig8}. However, the model's performance using a single frequency band is notably lower than when using all five frequency bands. 
\\

The experimental results demonstrate the effectiveness of our proposed method across different datasets. Despite facing challenges due to significant distributional disparities between the training and test data, our method performs significantly better compared to other cutting-edge methods.

\section{Conclusion}
\label{sec:conclusion}
This study proposes an efficient, unsupervised deep domain adaptation approach for recognizing emotions from EEG signals, addressing challenges such as limited labeled training data and differences in data distributions among datasets. Our proposal introduces the Gradual Proximity-guided Target Data Selection (GPTDS) technique, which gradually selects reliable target domain samples for training by considering their proximity to the source clusters and the model's confidence in predicting them. This approach prevents negative transfer resulting from the inclusion of diverse and unreliable samples from the target domain during training. We also propose a cost-effective test-time augmentation technique called Prediction Confidence-aware Test-Time Augmentation (PC-TTA). This technique applies TTA when necessary to improve the model's performance on test data while minimizing the computational burden posed by traditional TTA approaches. The experimental results across datasets demonstrate that the proposed approach yields more reasonable results in emotion recognition without using target domain labels during training, compared to existing state-of-the-art methods. The proposed approach has substantial industrial potential, offering enhanced human-computer interaction through more empathetic AI responses and providing robust solutions for monitoring and managing mood disorders in healthcare. Its adaptability across different platforms and reduced computational costs make it highly suitable for diverse, resource-constrained environments.

Although our method demonstrates significant improvements, it may struggle with extreme dataset variations and challenging cross-dataset adaptation scenarios. Future work will focus on generalizing our model to support multiple target domains by incorporating advanced domain alignment techniques to handle diverse and heterogeneous datasets. We will integrate functional connectivity between EEG electrodes by developing methods to capture and leverage dynamic relationships between brain regions. Additionally, we will address overfitting through advanced data augmentation \cite{kumar2023brain} and regularization \cite{haq2023dcnnbt,yousef2023u} techniques, including dropout and dynamic learning rate adjustments. While our PC-TTA method enhances computational efficiency and reduces the need for extensive test-time augmentations, we still need to evaluate its real-time performance and latency to ensure its suitability for latency-sensitive applications.

\section{Acknowledgments}
\label{sec:acknowledgment}
The authors express their gratitude to the Natural Sciences and Engineering Research Council of Canada (NSERC) (Grant No. NSERC RGPIN-2020-05471) and the New Frontiers in Research Fund (NFRF) (Grant No. NFRF 2021-00343) for their financial support.

\end{document}